\documentclass[doublecol]{epl2} 
\usepackage{amsmath}
\usepackage{amssymb}
\newcommand{\bX}{\textbf{X}}
\newcommand{\bx}{\textbf{x}}
\newcommand{\comment}[1]{}
\newcommand{\bv}{\mathbf{v}}
\newcommand{\mi}{\mathcal{I}}
\newcommand{\bF}{\mathbf{F}}
\newcommand{\bEta}{\boldsymbol{\eta}}

\newcommand{\ebX}{\epsilon {\bf X}}

\title{Irreversible effects of memory}
\shorttitle{Irreversible effects of memory} %Insert here a short version of the title if it exceeds 70 characters

\author{A. Puglisi\inst{1,2} \and D. Villamaina\inst{1,2}}
\shortauthor{A. Puglisi and D. Villamaina}

\institute{                    
  \inst{1} CNR-INFM SMC and ISC - p.le A. Moro 2, 00185, Roma, Italy\\
  \inst{2} Dipartimento di Fisica, Universit\`a Sapienza, p.le A. Moro 2, 00185, Roma, Italy
}
\pacs{05.40.-a}{Fluctuation phenomena, random processes, noise, and Brownian motion}
\pacs{05.70.Ln}{Nonequilibrium and irreversible thermodynamics}
\abstract{The steady state of a Langevin equation with short ranged
  memory and coloured noise is analyzed. When the
  fluctuation-dissipation theorem of second kind is not satisfied, the
  dynamics is irreversible, i.e. detailed balance is violated. We show
  that the entropy production rate for this system should include the
  power injected by ``memory forces''. With this additional
  contribution, the Fluctuation Relation is fairly verified in
  simulations. Both dynamics with inertia and overdamped dynamics
  yield the same expression for this additional power. The role of
  ``memory forces'' within the fluctuation-dissipation relation of
  first kind is also discussed. }

\begin{document}

\maketitle

\section{Introduction}

Irreversibility of path probabilities accompanies the lack of
thermodynamic equilibrium. This fact, which is obvious in relaxing
systems, can also be shown in many models of statistically
steady states. In the latter case there is the evidence for a general
connection between irreversibility and power dissipated by
non-conservative forces. A paradigm of this situation is a particle
performing Brownian motion under the action of conservative and
non-conservative forces:
\begin{equation}
\dot{\bv}=-\gamma \bv-\nabla U(\bx)+\bF_{nc}[\bx(t)]+\bEta,
\end{equation}
with $\bEta$ a Gaussian white noise, $\langle \bEta \rangle=0$ and
$\langle \eta_i(t) \eta_j(t') \rangle=2\gamma
T\delta_{ij}\delta(t-t')$, where $i,j$ are component
indices.  It is easily shown that
\begin{equation} \label{work}
W_t=\log\frac{P(\{\bv(s)\}_0^t)}{P(\{\mi\bv(s)\}_0^t)} \approx \frac{1}{T}\int_0^t \bF_{nc}[\bx(s)] \cdot \bv(s) ds.
\end{equation}
where $\{\mi\bv(s)\}_0^t$ is the time-reversed trajectory.
$W_t$ is usually called ``produced
entropy''~\cite{K98,LS99,seifert05}, and in some deterministic models
with certain assumptions it is equivalent, neglecting finite
differences, to the time-integral of the phase space contraction rate~\cite{ECM,ES94,GC}.

A phenomenon which may be  enhanced in statistically steady
states under the effect of non-conservative forces and/or multiple
baths is that of correlations among degrees of freedom: if
one insists on a reduced description of the system, such as the motion
of a tracer particle, these correlations must be reintroduced as
retarded feedback, or memory, accompanied by colored noise:
\begin{align}
-\gamma \bv &\to -\int_{-\infty}^t \gamma(t-t') \bv(t')dt'\\
\langle \eta_i(t) \eta_j(t') \rangle &\to M_{ij}(t-t').
\end{align}
We stress the fact that memory acts in equilibrated systems too,
as extensively discussed in classical references~\cite{KTH91},
where the condition for thermal equilibrium, in the absence of
non-conservative forces, is given by the fluctuation-dissipation
relation of the second kind
\begin{equation} \label{fdt2}
M_{ij}(t-t')=T\delta_{ij}\gamma(|t-t'|),
\end{equation}
which guarantees: (a) canonical distribution, (b) equipartition
between the particle and the surrounding fluid, (c) the validity of
the fluctuation-dissipation relation (FDR) of the first
kind\footnote{where $\overline{\cdot}$ and $\langle \cdot \rangle$ denote perturbed and unperturbed averages, respectively.}~\cite{BPRV08}
\begin{equation} \label{fdt1}
\frac{\overline{\delta v(t)}}{\delta v(0)}=\frac{1}{T}\langle v(t)v(0)\rangle.
\end{equation}
Violations of Eq.~\eqref{fdt1} or~\eqref{fdt2} are associated with lack
of equilibrium, i.e. presence of more than one thermostat and/or of
non-conservative external forces. This is, for instance, the case of a
granular liquid, that is a gas of inelastic hard particles coupled to
an external thermostat and at a packing fraction between $20\%$ and
$50\%$: in this system it appears that particles are correlated among
each others and this is reflected in a non-Markovian dynamics for a
tracer, together with a violation of the Einstein
relation~\cite{PBV07,VPV08,VBPV09}. A similar case is realized in a
molecular liquid coupled to different thermostats, where the motion of
a massive tracer is described by a Langevin equation with (short
ranged) memory, and Eq.~\eqref{fdt2} is not satisfied~\cite{CK00}.

Entropy production for Langevin models with short-ranged memory has
been recently studied in~\cite{SS07,OO07,MD07,ZBCK05}. In the first
three papers~\cite{SS07,OO07,MD07}, the steady state is treated under
the assumption of validity of~\eqref{fdt2}, called ``equilibrium''
bath,
%\footnote{It is not clear how realistic is
%Eq.~\eqref{fdt2} in the presence of external non-conservative forces
%$\bF_{nc}$, as proposed in many references, eg
%in~\cite{ZBCK05,MD07,OO07,SS07}: indeed, the appearance of a memory
%kernel and coloured noise follows projection/truncation of
%non-linear microscopic equations, and non-conservative forces
%should be incorporated from the beginning. In such a case, the noise
%covariance $M_{ij}(t)$ could depend upon $\bF_{nc}$ and
%Eq.~\eqref{fdt2} could be violated even in the presence of a single
%external thermostat.}
so that the irreversibility takes again the form of work done by
external non-conservative forces, similar to
Eq.~\eqref{work}. In~\cite{OO07} a non-transparent general formula,
for the case where~\eqref{fdt2} is violated, is also given, involving
a triple time-convolution between forces, memory $\gamma(t)$ and the
inverse of noise color $M(t)$. In~\cite{ZBCK05}, irreversibility of
paths is given as a convolution of power dissipated by all
(non-conservative and conservative) forces with a time-dependent
``effective temperature'' $T_{eff}(t)$ which characterizes the
violation of~\eqref{fdt2}. In the formulation of~\cite{ZBCK05}, the
entropy production vanishes if both anharmonic potential and 
non-conservative external forces are absent.

In this letter we show, for rapidly decaying memory kernels with
different time-scales, within a class of possible violations of
Eq.~\eqref{fdt2}, a simple formula, Eq.~\eqref{memory_ep}, for the
entropy production: in our formula memory appears as a force
performing work on the system, if and only if relation~\eqref{fdt2} is
not satisfied. The fact that memory alone produces entropy has not
been explicitly discussed in other references: this discrepancy is
likely to be due to a different definition for entropy production. We
consider the one given by Lebowitz and Spohn~\cite{LS99}, which is
related to the time-derivative of the Gibbs Entropy and which is
properly derived if the dynamics is Markovian, i.e. all degrees of
freedom required to determine the future are known in the present. To
this purpose we consider a class of non-Markovian dynamics which can
be mapped to Markovian systems by introducing auxiliary degrees of
freedom. An analogous approach is adopted in \cite{VBPV09} in order to
give an interpretation of ``violations" of FDR.

It could be argued that our formula includes contributions to the
entropy production coming from ``internal forces'', which is not
physical in a steady state. On the contrary, we stress the general
fact that memory cannot exist for an isolated particle: stated
differently, the effect of past history influences the present only if
other degrees of freedom (left out from the description) ``remind'' it
to the particle. Recollisions in dense fluids constitute a typical
mechanism~\cite{m89}. Therefore memory should be seen as an
``external'' but hidden force.  In the models considered here,
Eqs.~\eqref{gle} and~\eqref{glex}, this role is made explicit when the
auxiliary variables are introduced. We stress the fact that, since
only steady states are investigated here, cases with long-range memory
are not analyzed~\cite{HT09}.

The plan of the paper is the following: we first discuss our general
formula, and give details for the inertial and overdamped dynamics of
a tracer, showing that the Fluctuation Relation~\cite{K98,LS99} is
verified for entropy production when the memory contribution is taken
into account; then we analyze the connection between memory forces and
the violation of FDR, previously studied
in~\cite{VBPV09}; finally we draw conclusions and perspectives.

%%%%%%%%%%%%%%%%%%%%%%%%%%%%%%%%%%%%%%%%%%
\section{Irreversibility of multivariate paths}

Let us first discuss a non-multiplicative multivariate Langevin equation
with $N$ degrees of freedom and {\em without} memory:
\begin{equation} \label{multi}
\dot{X_{i}}=D_{i}(\bX)+g_{ij}\xi_j(t),
\end{equation}
with $i \in [0,N-1]$, $\xi_i(t)$ is a Gaussian process with
$\left<\xi_{i}(t)\right>=0$ and $\left<\xi_{i}(t)\xi_{i}(t')\right>=
2\delta_{ij}\delta(t-t')$. The probability distribution $f_t(\bX)$
satisfies
\begin{equation}
\frac{\partial f_t(\bX)}{\partial t}=-\sum_i\frac{\partial S_i(\bX)}{\partial X_i}
\end{equation}
with the probability current defined by
\begin{equation}
S_i(\bX)=D_i(\bX) f_t(\bX)-\sum_j\frac{\partial}{\partial X_j}D_{ij}f_t(\bX)
\end{equation}
where $D_{ij}= \sum_k g_{ik} g_{jk}$ is symmetric by construction.

Variables $X_i$ are assumed to have a well-defined parity $\epsilon_i=\pm 1$, with respect
to time-reversal. This leads to recognize reversible and irreversible
parts of the drift:
\begin{equation} \label{split}
D_i(\bX)=D_i^{rev}(\bX)+D_i^{ir}(\bX)
\end{equation}
with
\begin{align} 
D_i^{rev}(\bX)&=\frac{1}{2}[D_i(\bX)-\epsilon_i D_i(\ebX)]=-\epsilon_iD_i^{rev}(\ebX) \label{drev}\\
D_i^{ir}(\bX)&=\frac{1}{2}[D_i(\bX)+\epsilon_i D_i(\ebX)]=\epsilon_iD_i^{ir}(\ebX) \label{dirr}
\end{align}
having defined $\ebX=(\epsilon_0 X_0, \epsilon_1 X_1,
... \epsilon_{N-1} X_{N-1})$. This decomposition can be extended to
the probability current:
\begin{equation}
S_i^{rev}=f_tD_i^{rev} \;\;\;\;\;\;\; S_i^{ir}=S_i-S_i^{rev}=f_tD_i^{ir}-\sum_j D_{ij}\frac{\partial f_t}{\partial X_j}. \label{currents}
\end{equation}
%Eqs.~\eqref{drev} and~\eqref{dirr} are coherent with the fact that,
%upon time reversal, $\partial /\partial t$ changes sign too. 
%It can be
%verified (see~\cite{R89}) that a necessary condition for detailed
%balance is $S_i^{ir}=0$ in the stationary state. 
%Therefore, if $D^{ir}_i=0$, a path and its time-reversal have the
%same stationary probability; this is not true if $D^{ir}_i\neq 0$.

Following classical references~\cite{OM53,R89}, the expression for
conditional path probability of trajectory $\{\bX(s)\}_0^t$,
is\footnote{We use the Ito convention for stochastic integrals. Note
that, with this convention, the Jacobian in the path probability is $1$~\cite{H75}.}
\begin{multline} \label{onsager}
\log P(\{\bX(s)\}_0^t)=-\frac{1}{4}\sum_{jk}\int_0^t ds D^{-1}_{jk}\{\dot{X}_j(s)-D_j[\bX(s)]\}\\ \times 
\{\dot{X}_k(s)-D_k[\bX(s)]\},
\end{multline}
where we have assumed that $D_{ij}^{-1}$ exists.

By using Eq.~\eqref{split} and a few passages we get
\begin{multline}
W_t=\log\frac{P(\{\bX(s)\}_0^t)}{P(\{\mi\bX(s)\}_0^t)} = -\frac{1}{2}\int_{0}^{t}ds D^{-1}_{jk} \times \\ 
\left\{a^{-}_{jk} \left[\dot{X}_{j}\dot{X}_{k}+D^{ir}_{j}D^{ir}_{k}+D^{rev}_{j}D^{rev}_{k}-2\dot{X}_{j}D^{rev}_{k} \right]\right .\\
\left .-2a^{+}_{jk}\left[ D^{ir}_{j}\dot{X}_{k}-D^{ir}_{j}D^{rev}_{k}  \right] \right\},\label{formulone}
\end{multline}
where we have introduced the following definitions:
\begin{equation}
 a^{-}_{jk}=\frac{1-\epsilon_{j}\epsilon_{k}}{2}, \;\;\;\;\;\;\;\;
a^{+}_{jk}=\frac{1+\epsilon_{j}\epsilon_{k}}{2}.
\end{equation}
%such that $ a^{-}_{jk}+ a^{+}_{jk}=1$,
%$a^{-}_{jk}a^{+}_{jk}=0$, $a^{-}_{kk}=0$ and
%$a^{+}_{kk}=1$, 
Eq.~\eqref{formulone} is strongly simplified in the case of a
diagonal\footnote{for the models presented
here, this is the only relevant case.} diffusion matrix $D_{ij}$,
obtaining:
\begin{equation} 
W_t = \sum_{k}D^{-1}_{kk}\int_{0}^{t}ds D^{ir}_{k}\left[ \dot{X}_{k}-D^{rev}_{k}\right] \label{formulone2}.
\end{equation}
We are not aware of any previous derivation of this formula in the
literature. It is not difficult to generalize it to multiplicative
processes (i.e. $D_{ij}$ dependent upon $\bX$).

We now recall that, in order to obtain the complete path probability
in the steady state, one has to multiply $P(\{\bX(s)\}_0^t)$ by
$f(\bX(0))$, where $f=\underset{t\to +\infty}{\lim} f_t$ is the stationary
probability distribution. It is therefore possible to compute a
different quantity
\begin{align}
W_t'=\log\frac{f[\bX(0)]P(\{\bX(s)\}_0^t)}{f[\ebX(t)]P(\{\mi\bX(s)\}_0^t)}=W_t+b_t\\
b_t=\log\{f[\bX(0)]\}-\log\{f[\ebX(t)]\}.
\end{align}
The term $b_t$ is usually known as ``border term''~\cite{PRV06}. 

It is interesting also to define ``entropy production rates'' $\sigma$
and $\sigma'$, such that $W_t=\int_0^t \sigma(s)ds$ and $W_t'=\int_0^t
\sigma'(s)ds$. The  following decomposition can be done:
\begin{align}
\sigma&=\frac{d}{dt}(\log f)+\sigma_1+\sigma_2, \;\;\;\;\;\;\;\; \sigma'=\sigma_1+\sigma_2\\
\sigma_1&=\sum_i\dot{X}_i\left(-\frac{\partial \log f}{\partial X_i}+D_{ii}^{-1}D_i^{ir} \right) \label{sigma1}\\
\sigma_2&=-\sum_i D_{ii}^{-1}D_i^{ir}D_i^{rev}.
\end{align}
This decomposition will be particularly useful in the last section,
when discussing the connection with the FDR.

The condition of detailed balance is equivalent to $\sigma' \equiv 0$
for all trajectories in the steady state. When detailed balance does
not hold, it can be shown that
\begin{equation} \label{fr}
\log \frac{\textrm{p}(W_t'=x)}{\textrm{p}(W_t'=-x)}=x,
\end{equation}
where $p(W_t'=x)$ is the probability in the steady
state. Eq.~\eqref{fr} is the finite-time Fluctuation Relation (FR). In
general, excluding cases discussed in the
literature~\cite{othertheory,ESR,PRV06,BGGZ}, for large times $t$ one
has $W_t \approx W_t'$ and relation~\eqref{fr} is also satisfied by
$W_t$~\cite{K98,LS99,seifert05}.

%%%%%%%%%%%%%%%%%%%%%%%%%%%%%%%%
\section{Memory through auxiliary variables}

The first class of generalized Langevin equations with memory we are
interested in, concerns inertial dynamics, i.e. we consider the
following equation of motion (we restrict ourselves to the one-dimensional problem,
without loss of generality):
\begin{equation} 
\left\{ \begin{array}{ccl}\dot{x}&=&v\\
\dot{v}&=&F(x)-\int^{t}_{-\infty}\gamma(t-t')v(t')dt'+\eta(t)\end{array} \right.\label{gle}
\end{equation}
with 
\begin{align}
\gamma(t)&=2\gamma_0 \delta(t)+\sum_{i=1}^M \frac{\gamma_i}{\tau_i} e^{-\frac{t}{\tau_i}} \;\;\;\;\;\;\;\;\; \langle \eta(t) \rangle=0 \label{friction}\\
\langle \eta(t)\eta(t') \rangle&=2T_0\gamma_0 \delta(t-t')+\sum_{i=1}^M T_i \frac{\gamma_i}{\tau_i} e^{-\frac{|t-t'|}{\tau_i}}. \label{noise}
\end{align}
and where $F$ is a generic drift term which can take the form of a sum
of conservative and non-conservative forces, i.e. $F(x)\equiv
-\frac{dU_0(x)}{dx}+F_{nc}(x)$.  When $T_i=T_0$ for all $i$, the FDR
of the second kind, Eq.~\eqref{fdt2}, holds. This model has
several applications: among others, it has been proposed
in~\cite{CK00} for weakly driven glassy systems; the dynamics of a
tracer particle in moderately dense fluidized granular media,
including its linear response properties, are consistent with this
model~\cite{PBV07}, and, recently, the noise in feedback cooled
oscillator for gravitational wave detectors~\cite{B09} has been
characterized in a similar
fashion~\cite{GRBC09}.\footnote{Note that the pairing of
equal characteristic times for the exponentials in~\eqref{friction}
and~\eqref{noise} is not so restrictive: indeed, case $T_i=0$ or case
$\gamma_i \to 0$, $T_i \to \infty$ with finite $\gamma_i T_i$, for
some $i$, can be easily worked out and make no exception to the
following analysis.}

It is useful to map Eq.~\eqref{gle} into Eq.~\eqref{multi}, where all
noises are uncorrelated, identifying $N=M+2$ and $X_0\equiv v$, $X_{N-1} \equiv x$, while
$X_i\equiv v_i$ ($i \in [1,M]$) are auxiliary variables necessary to
take into account memory, for instance they can be defined as
\begin{equation} \label{aux}
v_i(t)=\sqrt{\frac{\gamma_i}{\tau_i}}\int_{-\infty}^t e^{-\frac{t-t'}{\tau_i}}\left(v(t')+\sqrt{\frac{T_i}{\gamma_i}}\xi_i(t') \right)dt'.
\end{equation}
With this choice for the auxiliary variables, it is easy to verify that the drifts in Eq.~\eqref{multi} are
\begin{align}
D_0&=F(x)-\gamma_0 v-\sum_{i=1}^M\sqrt{\frac{\gamma_i}{\tau_i}}v_i \\
D_i&=\sqrt{\frac{\gamma_i}{\tau_i}}v-\frac{1}{\tau_i}v_i \;\;\;\;(i \in [1,M])\\
D_{N-1}&=v,
\end{align}
while the diagonal diffusion matrix reads
%\begin{equation}
%D=\mathbb{I}\begin{pmatrix}\gamma_0T_0 \\\frac{T_1}{\tau_1}\\\vdots\\\frac{T_M}{\tau_M}\\0\end{pmatrix}
%\end{equation}
\begin{equation}
D_{00}=\gamma_0T_0 \;\;\;\; D_{ii}=\frac{T_i}{\tau_i}
\end{equation}
%where $\mathbb{I}$ is the $N\times N$ identity matrix.
Summarizing, the system with memory is recast into a system of
(linearly) coupled Langevin equations where all noises are
uncorrelated.
Auxiliary variables $v_i$ ($i \in [1,M]$) are even under
time-reversal, i.e. $\epsilon_i=1$ for $i>0$: this can be understood,
for instance, requiring the validity of detail balance in the
equilibrium case $T_i=T_0$ for all $i$.

%, we use the following argument: since the
%explicit application of time-reversal to $v(s)$, gives
% $\mi v(s)=\epsilon_0
%v(t-s)$,\footnote{in this case $X_0 \equiv v$, i.e.
%$\epsilon_0=-1$.} for the same realization of noise, the only
%possibility for $\{\mi v(s)\}_0^t$ to be a solution of
%Eq.~\eqref{gle} is to have $\mi v_i(0)=v_i(t)$,
%i.e. $\epsilon_i=1$ for $i>0$. 

Then we obtain
\begin{align}
D_0^{rev}&=F-\sum_{i=1}^M\sqrt{\frac{\gamma_i}{\tau_i}}v_i, \;\;\;\; D_0^{ir}=-\gamma_0 v  \\
D_i^{rev}&=\sqrt{\frac{\gamma_i}{\tau_i}}v, \;\;\;\; D_i^{ir}=-\frac{v_i}{\tau_i}  \;\;\;\;\;(i \in [1,M])\\
D_{N-1}^{rev}&=v, \;\;\;\;\;\;\;\;\;D_{N-1}^{ir}=0. 
\end{align}
%The interesting fact is that $D_i^{ir} \propto v_i$, meaning that
%$\int_0^t D_i^{ir}(s) \dot{v}_i(s) ds\propto
%\frac{1}{2}[v_i^2(t)-v_i^2(0)]$.  In fact, from Eq.~\eqref{formulone2}, one gets

When computing Eq.~\eqref{formulone2}, it is crucial to note that
$D_{ij}$ is not positive definite and cannot be inverted. Anyway, as
noted by Machlup and Onsager~\cite{MO53}, the last row and column of
$D_{ij}$ (those associated to variable $x$, which has not explicit
noise dependence) can be dropped out for the purpose of computing path
probabilities. With this observation, formula~\eqref{formulone2} can
be used, leading to
\begin{align} \label{memory_ep}
\begin{split}W_t&=-\sum_{i=0}^M\frac{\delta(v_i^2)}{2T_i}-\frac{\delta U_0}{T_0}+
\\&+\int_0^t \frac{1}{T_0}\left(F_{nc}[x(s)]+\sum_i F_i[v_i(s)] \right)v(s)ds\end{split}\\
F_i&=-\sqrt{\frac{\gamma_i}{\tau_i}}\left(1-\frac{T_0}{T_i} \right) v_i(s). \label{memfor}
\end{align}
As usual, exact differences appear, denoted as $\delta(g)\equiv g(t)-g(0)$. The
non-trivial part of $W_t$ is the time-integral on the
r.h.s. of~\eqref{memory_ep}, which does not reduce to exact
differences: it is equivalent to the work done by the usual
non-conservative external force $F_{nc}(x)$ and by new forces $F_i$
expressed in Eq.~\eqref{memfor}.  It can be verified that memory
forces do not depend on the definition of auxiliary variables,
Eq.~\eqref{aux}, as expected.  The additional work done by forces
$F_i$ is due to feedback of past history on the particle velocity and
it is interesting to discover its effect on irreversibility. From
formula~\eqref{memory_ep} it is also evident that force $F_i$ vanishes
if $T_i=T_0$. If $T_i=T_0$ for all $i$, then the second kind FDR,
Eq.~\eqref{fdt2}, holds, and memory does not contribute to $W_t$.

We conclude this section, evaluating the so-called finite time (or transient)
contribution $b_t$ which must be added in order to verify the
FR at short times~\cite{PRV06}, in the
particular case $M=1$ (only one auxiliary variable). We consider a setup where there is no external force, $F=0$, apart from two
reflecting walls confining the particle to have a uniform spatial
distribution in between, making irrelevant the contribution of $x$ to the stationary probability density.
Considering only variables $\bv = (v_0 \equiv v,v_1)$, one has:
%\frac{[\mathrm{Det} \Sigma^{-1}]^{1/2}}{(2\pi)^{N/2}}
\begin{equation}
f(\bv) \propto\exp\left(-\frac{1}{2}\sum_{ij}\Sigma^{-1}_{ij}v_i v_j \right),
\end{equation}
where the inverse covariance matrix elements read
$\Sigma^{-1}_{00}=\frac{1}{T_0}\left(1+\frac{T^*}{Q}\Delta T \right)$,
$\Sigma^{-1}_{11}=\frac{1}{T_1}\left(1-\frac{\gamma_0 \tau_1 T^*}{Q}
\Delta T\right)$ and
$\Sigma^{-1}_{01}=-\frac{\gamma_0}{\gamma_1}\sqrt{\gamma_1 \tau_1}\frac{1+\gamma_0\tau_1}{Q}\Delta
T$, with $\Delta T=T_0-T_1$, $T^*=\gamma_0\tau_1T_0+T_1$ and
$Q=(T^*)^2+\frac{\gamma_0}{\gamma_1}T_0T_1(1+\gamma_0\tau_1)^2$.
In summary:
\begin{equation}
b_t=\Sigma^{-1}_{00}\frac{\delta{v^{2}}}{2}+\Sigma^{-1}_{11}\frac{\delta{v_1^{2}}}{2}-\Sigma^{-1}_{01}[v_1(0)v(0)+v_1(t)v(t)] \label{term3}
\end{equation}

Let us note that, when $T_0=T_1$ (validity of the
FDR of the second kind), the exact
difference (first term) of $W_t$ exactly cancels $b_t$: in particular,
if external non-conservative forces are absent, it appears that
$W_t'=0$, i.e. detailed balance is satisfied.

\begin{figure}
\includegraphics[width=8.5cm,clip=true]{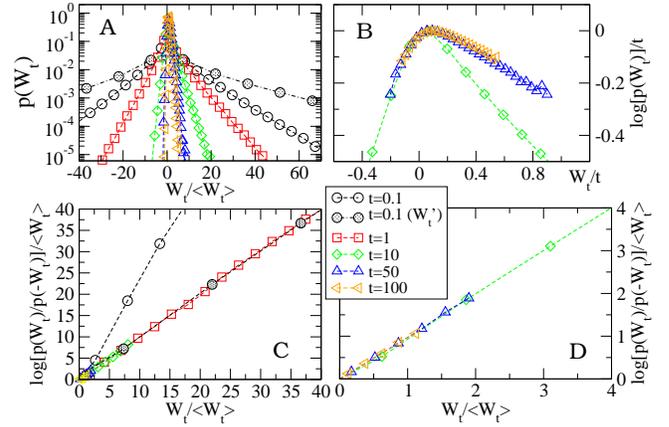}
\caption{A: pdf of $W_t$ for the inertial Langevin
equation~\eqref{gle} with simple exponential memory, i.e. $M=1$,
$T_0=0.6$, $T_1=3$, $\gamma_0=10$, $\tau_1=10$, $\gamma_1=5$, and
$F=0$, for different times $t$ of integration. B: at large times,
$\log p(W_t/t)/t$ converges to a time-independent function: the large
deviation rate. C-D: check of the
Fluctuation Relation (FR), Eq.~\eqref{fr} which is verified for all
data aligned along the bisector. At small time (empty circles), where
the FR does not hold, Eq.~\eqref{fr} is verified for the pdf of $W_t'$
(gray circles). Note that the only source of entropy, for this
example, is memory.\label{fig:under}}
\end{figure}

In Figure~\ref{fig:under} we show the probability density function
(pdf) $p(W_t)$, in the steady state, obtained numerically by
integrating Eq.~\eqref{gle}, for different choices of times. In the
same figure we also show the validity of Eq.~\eqref{fr} for $W_t$ at
large times and $W_t'$ at any time (see the difference between empty
and gray circles), as well as the asymptotic convergence to the large
deviation rate function.

%%%%%%%%%%%%%%%%%%%%%%%%%%%%%%%%%%%%%%%%%%%%%%%%%%%%%%%%%%%%%%%%%%%%%%%%
\section{The  overdamped limit}

In the overdamped limit, the role of main variable is played by the
tracer position $x$, while velocity $v$ is neglected. Even if the
dynamics is changed, and drift terms have different
symmetries with respect to time-reversal, the final result appears
identical, making robust our observation.

The overdamped limit of Eq.~\eqref{gle} is a generalized Langevin
equation with memory for the tracer position:
\begin{multline} 
\gamma_0\dot{x}=-\frac{dU_0(x)}{dx}+F_{nc}(x)+\\
-\sum_{i=1}^M\frac{\gamma_i}{\tau_i}\int^{t}_{-\infty} e^{-\frac{t-t'}{\tau_i}}\dot{x}(t')dt'+\eta(t)\label{glex}
\end{multline}
with the same properties as in Eq.~\eqref{noise} for the noise $\eta(t)$.
%\begin{equation}
%\langle \eta(t)\eta(t') \rangle=2T_0\gamma_0 \delta(t-t')+\sum_{i=1}^M T_i \frac{\gamma_i}{\tau_i} e^{-\frac{t-t'}{\tau_i}}.
%\end{equation}
With a partial integration it is possible to cast Eq.~\eqref{glex} into
\begin{multline} 
\gamma_0\dot{x}=-\frac{dU_0(x)}{dx}+F_{nc}(x)-\sum_{i=1}^M\frac{\gamma_i}{\tau_i}x+\\+\sum_{i=1}^M\frac{\gamma_i}{\tau_i^2}\int^{t}_{-\infty}  e^{-\frac{t-t'}{\tau_i}}x(t')dt'+\eta(t)\label{glex2}.
\end{multline}
We use now the following auxiliary variables, with $i \in [1,M]$:
\begin{equation} \label{aux2}
x_i(t) = \frac 1{\tau_i}\int_{-\infty}^t
e^{-\frac{t-t'}{\tau_i}}\left(x(t')+\tau_i\sqrt{\frac{T_i}{\gamma_i}}\xi_i(t')\right)dt',
\end{equation}
obtaining an equivalence with a system of Langevin equations without
memory of the form~\eqref{multi}, with $N=M+1$ (identifying $X_0
\equiv x$ and $X_i \equiv x_i$), and with
\begin{align}
\gamma_0 D_0&=-\frac{dU_0(x)}{dx}+F_{nc}(x) -\sum_{i=1}^M\frac{\gamma_i}{\tau_i}(x-x_i)\\
D_i&=\frac{1}{\tau_i}(x-x_i) \;\;\;(i>0), \;\;\;\;\;\;\;\; D_{ij}=\delta_{ij}\frac{T_i}{\gamma_i}.
\end{align}

Following the same procedure of identification of irreversible and
reversible parts of drifts, and recognizing again that memory
auxiliary variables have parity $\epsilon_i=1$, we obtain
\begin{equation}
D_i^{rev}=0, \;\;\;\;\; D_i^{ir}=D_i.
\end{equation}
This leads to identifying, for the entropy production, the following
expression, after suitable partial integrations:
\begin{multline} \label{memory_ep2}
W_t=-\sum_{i=0}^M\frac{\delta U_i}{T_i}+\int \frac{1}{T_0}\left \{ F_{nc}[x(s)] +\phantom{\sum_{i=1}^M}\right.\\
\left. +\sum_{i=1}^MF_i[x(s)] \right\}\dot{x}(s)ds
\end{multline}
where
\begin{align}
U_i(t)&=\frac{\gamma_i}{2\tau_i}[x(t)-x_i(t)]^2 \;\;\;\;\;(i>0)\\
F_i&=\frac{\gamma_i}{\tau_i}\left(1-\frac{T_0}{T_i}\right)[x_i(s)-x(s)] \label{memfor2}.
\end{align}
As seen, also for the overdamped case, entropy production is
equivalent to the work done by the external non-conservative force
plus non-conservative forces $F_i(s)$ due to memory. It is easy to verify,
through a partial integration and the comparison between Eq.~\eqref{aux}
and Eq.~\eqref{aux2}, that 
\begin{equation}
\frac{\gamma_i}{\tau_i}[x_i(t)-x(t)]=-\sqrt{\frac{\gamma_i}{\tau_i}}v_i(t),
\end{equation}
and therefore the forces in Eq.~\eqref{memfor2} are exactly equivalent
to the forces in Eq.~\eqref{memfor} for the inertial dynamics.
Again we have verified, through numerical integration of
Eq.~\eqref{glex}, that the pdf of $W_t$ for large times reproduces the
FR. For short times, the terms $b_t$ due to the
steady state invariant measure at initial and final configurations
must be added in order to recover the short-times kind of symmetry. The
same consideration drawn for the inertial dynamics can be repeated
here: when $T_i=T$, for all $i$, i.e. when the FDR of the second kind holds, then $F_i=0$ and the boundary term
$b_t$ cancels out the exact differences in Eq.~\eqref{memory_ep2}, so
that, if $F_{nc}=0$, detailed balance holds. 

In conclusion, the only difference between
expression~\eqref{memory_ep} and~\eqref{memory_ep2} for the entropy
production, is given by exact differences: these
differences, for large times, can be neglected and the two expressions
become equivalent. This is coherent with the fact that the overdamped
dynamics ignores the short time-scale corresponding to the
relaxation of velocity.

%%%%%%%%%%%%%%%%%%%%%%%%%%%%%
\section{Linear response}

The first-kind FDR for the overdamped model in~\eqref{glex} has been
treated in several papers, see for instance~\cite{CK00,ZBCK05}. More
in general, it is known that a generalized
FDR~\cite{A72,DH75,FIV90,BPRV08,VBPV09} is satisfied if all variables
$\bX\equiv{X_0,X_1,\dotsc,X_M}$ are taken into account with their
steady state invariant measure $f(\bX)$, provided that
it is smooth and non vanishing, and the system is mixing.

If an impulsive variation of coordinates $\{\delta X_i(0)\}$
at time $0$ is considered, the generalized FDR for the response  takes the form
\begin{align} \label{gfdr}
R_{ji}&\overset{def}{=}\frac{\overline{\delta X_j(t)}}{\delta X_i(0)}=\left\langle X_j(t) B_i(0)\right\rangle\\
B_i&=-\frac{\partial \log f(\bX)}{\partial X_i},
\end{align}
where we use $\overline{\cdot}$ to mean non-equilibrium averages
following the perturbation at time $0$, while $\langle \cdot \rangle$
represents an ensemble average, which (under ergodicity) is equivalent
to averaging over a long trajectory in the stationary
state.  If the
diffusion matrix, $D_{ij}$, is diagonal, one has~\cite{R89}:
\begin{align}
B_i&=B_i^0+B_i^*\\
B_i^0&=-D_{ii}^{-1}D_i^{ir}, \;\;\;\;\;\;\;B_i^*=\frac{S_i^{ir}}{D_{ii}f}. \label{bistar}
\end{align}
%From Eq.~\eqref{gfdr}, we obtain
%\begin{equation} \label{gfdr2}
%R_{ji}=R_{ji}^0+R_{ji}^*, \;\;\;\;\;\;\; R_{ji}^{0(*)}=\langle x_j(t) B_i^{0(*)}(0)\rangle\\
%\end{equation}

When detailed balance is satisfied, which implies that $W_t=0$ on
average and $W_t'=0$ for each trajectory, one has $S_i^{ir}=0$~\cite{R89} and therefore
\begin{equation} \label{fdt0}
R_{ji}=R_{ji}^0\overset{def}{=}\langle X_j(t) B_i^0(0)\rangle=-\left\langle X_j(t) \frac{D_i^{ir}(0)}{D_{ii}}\right\rangle.
\end{equation}
It is straightforward
to verify that Eq.~\eqref{fdt0} takes standard equilibrium forms, e.g. it is
equivalent to Eq.~\eqref{fdt1} for velocity, or more in general to
Kubo relations~\cite{KTH91,R89}, which depends upon the choice of system,
perturbation and measured responses.
%: for instance, the response of
%variable $x$ in model~\eqref{glex} after the application of an
%impulsive force $F$ (corresponding to an impulsive displacement $\delta x(0)=F/\gamma_0$), reads
%\begin{equation}
%R_{xF}=\frac{\overline{\delta x(t)}}{\gamma_0\delta x(0)}=\frac{1}{T_0}\frac{d}{dt}\langle x(t)x(0)\rangle
%\end{equation}
%{\bf problema: manca un segno meno...}
As discussed in previous sections, for models~\eqref{gle}
and~\eqref{glex}, when external non-conservative
forces are absent, detailed balance condition $S_i^{ir}=0$ corresponds
to $T_i=T_0$ for all $i$ , i.e. to the FDR
of the second kind, Eq.~\eqref{fdt2}.

On the contrary, when detailed balance is not satisfied, the response includes an additive contribution:
\begin{equation}
R_{ji}=R_{ji}^0+\langle X_j(t) B_i^*(0)\rangle.
\end{equation}

Interestingly, comparison of Eqs.~\eqref{currents},~\eqref{sigma1}
and~\eqref{bistar} gives the following identification:
\begin{equation} \label{connect}
\sigma_1=\sum_i\dot{X}_iB_i^*.
\end{equation}
This relation illustrates the connection between the ``violation'' of
Eq.~\eqref{fdt0} and a part of the entropy production,
$\sigma_1$. Indeed, relation~\eqref{connect} becomes very useful in
the overdamped dynamics, where $\sigma_2=0$: in this case the
generalized force $B_i^*$ which entirely contributes to the entropy
production $\sigma'=\sigma_1$, is the same force acting as ``conjugate
quantity'' in the additional (non-equilibrium) terms of the linear response formula. In the case with inertia,
relation~\eqref{connect} becomes less useful, since $\sigma_1$ only
contains exact differential and the bulk contribution to the entropy
production comes from $\sigma_2$. 
%Nevertheless, as studied
%in~\cite{VBPV09}, when detailed balance is violated, $D_i^{ir}$ additional pieces appear in 
%the additive terms which must be added to the
%equilibrium formula in order to obtain the complete non-equilibrium
%formula are proportional to $\langle v(t)v_i(0) \rangle$, suggesting
%that a similar identification can be done.

%In~\cite{VBPV09} model~\eqref{glex}, as well as its underdamped
%counterpart, Eq.~\eqref{gle}, have been treated in the linear
%case. 

We wish to point out that other approaches toward the connection
between FDR and entropy production have also been discussed, from
different points of view, in many recent works,
e.g.~\cite{HS01,LCZ05,ss06,BMW09}: for a detailed review, see also~\cite{SS09}.

%%%%%%%%%%%%%%%%%%%%%%%%%%%%%%%%%%%%%%%%%%%%%%%%%%
\section{Conclusions}

%We have discussed the irreversible effects, in terms of entropy
%produced during a trajectory of timelength $t$, for a multivariate
%Langevin process: this formula is exploited to express in a simple
%form the entropy produced by a mono-variate Langevin process with
%short-ranged memory. We have shown that memory acts as a force which
%produces entropy, when detailed balance is globally not
%satisfied. 
Summarizing, we have discussed the role of memory in non-equilibrium
steady states, merging two main observations: 1) a general formula for
entropy production in multivariate memory-less Langevin processes and
2) the mapping between Langevin models with short-ranged memory toward
memory-less Langevin models with auxiliary variables. The latter 
leads to identify memory as a non-conservative force. These forces
cease to contribute to entropy production only under the validity of
the FDR of the second kind, Eq.~\eqref{fdt2}.

Interestingly, in the overdamped dynamics, these same forces
contribute to the so-called ``violations'' of the FDR of the first
kind, when detailed balance is not satisfied. Future work should
include a generalization to Langevin equations with other forms of
memory. Investigation of the case with inertia is also needed, to
better explore the connections with the linear response theory.

%%%%%%%%%%%%%%%%%%%%%

\acknowledgments The work of the authors is supported by the
``Granular-Chaos'' project, funded by the Italian MIUR under the
FIRB-IDEAS grant number RBID08Z9JE. Both authors wish to thank P. De
Gregorio, G. Gonnella, L. Rondoni, P. Visco and A. Vulpiani for useful
discussions.

\bibliographystyle{eplbib}
\bibliography{fluct.bib}

\begin{thebibliography}{10}
\expandafter\ifx\csname url\endcsname\relax\def\url#1{\texttt{#1}}\fi

\bibitem{K98}
\Name{Kurchan J.} \REVIEW{J. Phys. A }{31}{1998}{3719}.

\bibitem{LS99}
\Name{Lebowitz J.~L. \and Spohn H.} \REVIEW{J. Stat. Phys. }{95}{1999}{333}.

\bibitem{seifert05}
\Name{Seifert U.} \REVIEW{Phys. Rev. Lett. }{95}{2005}{040602}.

\bibitem{ECM}
\Name{Evans D.~J., Cohen E. G.~D. \and Morriss G.~P.} \REVIEW{Phys. Rev. Lett.
  }{71}{1993}{2401}.

\bibitem{ES94}
\Name{Evans D.~J. \and Searles D.~J.} \REVIEW{Phys. Rev. E }{50}{1994}{1645}.

\bibitem{GC}
\Name{Gallavotti G. \and Cohen E. G.~D.} \REVIEW{J. Stat. Phys.
  }{80}{1995}{931}.

\bibitem{KTH91}
\Name{Kubo R., Toda M. \and Hashitsume N.} \Book{Statistical physics II:
  Nonequilibrium stastical mechanics} (Springer) 1991.

\bibitem{BPRV08}
\Name{Marconi U. M.~B., Puglisi A., Rondoni L. \and Vulpiani A.} \REVIEW{Phys.
  Rep. }{461}{2008}{111}.

\bibitem{PBV07}
\Name{Puglisi A., Baldassarri A. \and Vulpiani A.} \REVIEW{J. Stat. Mech.
  }{}{2007}{P08016}.

\bibitem{VPV08}
\Name{Villamaina D., Puglisi A. \and Vulpiani A.} \REVIEW{J. Stat. Mech.
  }{}{2008}{L10001}.

\bibitem{VBPV09}
\Name{Villamaina D., Baldassarri A., Puglisi A. \and Vulpiani A.} \REVIEW{J.
  Stat. Mech. }{}{2009}{P07024}.

\bibitem{CK00}
\Name{Cugliandolo L.~F. \and Kurchan J.} \REVIEW{J Phys Soc Jpn
  }{69}{2000}{247}.

\bibitem{SS07}
\Name{Speck T. \and Seifert U.} \REVIEW{J. Stat. Mech. }{}{2007}{L09002}.

\bibitem{OO07}
\Name{Ohkuma T. \and Ohta T.} \REVIEW{J. Stat. Mech. }{}{2007}{P10010}.

\bibitem{MD07}
\Name{Mai T. \and Dhar A.} \REVIEW{Phys. Rev. E }{75}{2007}{061101}.

\bibitem{ZBCK05}
\Name{Zamponi F., Bonetto F., Cugliandolo L.~F. \and Kurchan J.} \REVIEW{J.
  Stat. Mech. }{}{2005}{P09013}.

\bibitem{m89}
\Name{McLennan J.~A.} \Book{Introduction to Nonequilibrium Statistical
  Mechanics} (Prentice-Hall) 1989.

\bibitem{HT09}
\Name{Harris R.~J. \and Touchette H.} \REVIEW{J. Phys. A: Math. Theor.
  }{42}{2009}{342001}.

\bibitem{OM53}
\Name{Onsager L. \and Machlup S.} \REVIEW{Phys. Rev. }{91}{1953}{1505}.

\bibitem{R89}
\Name{Risken H.} \Book{The Fokker-Planck equation: Methods of solution and
  applications} (Springer- {V}erlag, Berlin) 1989.

\bibitem{H75}
\Name{H\"anggi P.} \Book{Path integral solution for nonlinear {G}eneralized
  {L}angevin equations} in proc. of \Book{Path Integrals for meV to MeV:
  Tutzing '92}, edited by \Name{Grabert H., Inomata A., Schulman L. \and Weiss
  U.} (World Scientific) 1993 p. 289.

\bibitem{PRV06}
\Name{Puglisi A., Rondoni L. \and Vulpiani A.} \REVIEW{J. Stat. Mech.
  }{}{2006}{P08010}.

\bibitem{othertheory}
\Name{van Zon R. \and Cohen E. G.~D.} \REVIEW{Phys. Rev. Lett.
  }{91}{2003}{110601}.

\bibitem{ESR}
\Name{Evans D., Searles D. \and Rondoni L.} \REVIEW{Phys. Rev. E
  }{71}{2005}{056120}.

\bibitem{BGGZ}
\Name{Bonetto F., Gallavotti G., Giuliani A. \and Zamponi F.} \REVIEW{J. Stat.
  Phys. }{123}{2006}{39}.

\bibitem{B09}
\Name{Bonaldi M. \and et~al.} \REVIEW{Phys. Rev. Lett. }{103}{2009}{010601}.

\bibitem{GRBC09}
\Name{Gregorio P.~D., Rondoni L., Bonaldi M. \and Conti L.}
  \REVIEW{arXiv:0907.4309 }{}{2009}{}.

\bibitem{MO53}
\Name{Machlup S. \and Onsager L.} \REVIEW{Phys. Rev. }{91}{1953}{1512}.

\bibitem{A72}
\Name{Agarwal G.~S.} \REVIEW{Z. Physik }{252}{1972}{25}.

\bibitem{DH75}
\Name{Deker U. \and Haake F.} \REVIEW{Phys. Rev. A }{11}{1975}{2043}.

\bibitem{FIV90}
\Name{Falcioni M., Isola S. \and Vulpiani A.} \REVIEW{Physics {L}etters {A}
  }{144}{1990}{341}.

\bibitem{HS01}
\Name{Hatano T. \and Sasa S.} \REVIEW{Phys. Rev. Lett. }{86}{2001}{3463}.

\bibitem{LCZ05}
\Name{Lippiello E., Corberi F. \and Zannetti M.} \REVIEW{Phys. Rev. E
  }{71}{2005}{036104}.

\bibitem{ss06}
\Name{Speck T. \and Seifert U.} \REVIEW{Europhys. Lett. }{74}{2006}{391}.

\bibitem{BMW09}
\Name{Baiesi M., Maes C. \and Wynants B.} \REVIEW{Phys. Rev. Lett.
  }{103}{2009}{010602}.

\bibitem{SS09}
\Name{Speck T. \and Seifert U.} \REVIEW{arXiv:0907.5478 }{}{2009}{}.

\end{thebibliography}

\end{document}